\documentclass[11pt]{article}
\UseRawInputEncoding
\usepackage{amsmath,amssymb,amsthm}
\usepackage{geometry}
\usepackage{hyperref}
\usepackage{cite}
\usepackage{bbm}
\usepackage{mathtools}

\title{\large\bf Note on the Pure State of Old Black Holes in Some Models of Asymptotically Flat Quantum Gravity}

\author{Tom Banks\\
NHETC and Department of Physics \\
Rutgers University, Piscataway, NJ 08854-8019\\
E-mail: \href{mailto:tibanks@ucsc.edu}{tibanks@ucsc.edu}
\\
\\
}

\date{}

\begin{document}
\maketitle

\begin{abstract} We analyze two old models of black holes to see whether they realize the AdS/CFT paradigm of ``multiple representations of bulk data by boundary states that differ only in the expectation value of exponentially complex operators" .  We find that the explanation of the firewall paradox in these models is much simpler and corresponds to the naive way entanglement is resolved when small constituents escape from a complex chaotic quantum system.  The real issue is that the quantum field theory (QFT) picture of the states near the boundary of a causal diamond is wrong.  The phrase in quotes above is correct in our models because most of the states associated with any causal diamond cannot be associated with ANY states in quantum field theory.  
 
\end{abstract}

\tableofcontents

% =========================================================
\section{Introduction}
% =========================================================
A general consensus has formed that the firewall paradox\cite{firewall} is resolved by the fact that there are states in quantum gravity, which resemble the Hartle-Hawking state of effective field theory (EFT) in a black hole background for the computation of all local operators that are not of exponential complexity $e^{S_{BH}}$ in the vicinity of the black hole horizon.  As far as the present author can tell, this claim is based on the proven fact that in AdS/CFT local operators deep in the bulk are represented by extremely complicated non-local operators in the boundary CFT where {\it all} quantum information really resides.  It's then rather plausible that there are multiple choices of boundary operators, all of which produce the same results for simple bulk correlators but different results for detailed quantum information. In effect this is like saying that the simple bulk operators are only calculating coarse grained hydrodynamic\footnote{Here one must interpret hydrodynamics broadly, to include quantized phonons and Landau fermi liquid excitations.} properties of the quantum state, which is consistent with scale radius duality.   It's important to realize that these claims do {\it not} amount to saying that the EFT picture of the states near the black hole horizon is correct, and is quite the contrary of that.  In the conclusions we will note\cite{pdk} that there are no states in EFT compatible with the properties hypothesized in this AdS/CFT resolution of the firewall paradox.

The asymptotic boundary of Minkowski space is null and the Quantum Error Correcting Code/Tensor Network\cite{swingleetal} picture of locality that has been developed for AdS space does not apply to it.  In this paper we will study the fate of old black holes in two classes of models of quantum gravity in asymptotically flat space.  Local physics emerges in two apparently quite different ways in those models, though ultimately they can both be viewed as versions of matrix quantum mechanics.  In both models, high entropy meta-stable resonances with the properties of black holes can be built using the tools of ordinary non-relativistic quantum mechanics.  Both are intimately connected to string theory and can be used to calculate a variety of perturbative gravitational scattering amplitudes.  

The first of these models is a variation\cite{lindil} on the $\hat{c} = 1$ string theory\cite{c=1}, which actually has excitations with the properties of black holes appearing in the semi-classical model of CGHS\cite{cghs}.  We review that model and the solution of the models of\cite{c=1}.  Among the lessons this model teaches are the dangers of relying too heavily on arguments ``to all orders in string perturbation theory".
We show that the entanglement in EFT between particles on two sides of the horizon has no counterpart in the model.  One can view the black hole as a self consistent meta-stable bound state of non-relativistic fermions and the Hawking radiation as individual fermions tunneling out of the potential well.  

The second model is the model of black holes in the (M)atrix model of 11 dimensional quantum supergravity\cite{bfss}\cite{bfssbh} either in eleven dimensional space or when compactified on $T^3$.  In the eleven dimensional model black holes\footnote{These are really black strings wrapped around the direction that becomes null in the $N\rightarrow\infty$ limit.}  are viewed in the light front frame as individual matrix blocks bound to a classical solution of the matrix equations of motion.  The energy and entropy and size of these systems matches the scaling laws for black holes.  When compactified on $T^3$ the model becomes maximally supersymmetric Yang-Mills theory and we can estimate the energy entropy and size by using the conformal symmetry of that model.  In both cases, Hawking decay is described as the liberation of small matrix blocks (which are identified as D0 branes in the Type IIA string theory origin of the model) from the bound state.  In\cite{matrixbhdecay} the Hawking decay rate was estimated in the 11 dimensional version of the model and agreed with semi-classical predictions up to an overall order 1 constant.  

In neither of these models do we see any evidence of the infinite cross horizon entanglement predicted by EFT.  The ``Hawking particles" are constituents of a non-relativistic bound state (a rather exotic one in the (M)atrix Theory example) which are liberated and propagate out to infinity.  They're of course entangled with {\it what the state of the bound state was at the moment they departed}.  By unitary evolution, and without intervention by an external system, they remain entangled, but the state with which they're entangled is changing rapidly in time.  Right after the Page time the state of the whole system has the form
\begin{equation} | \psi \rangle = \sum a_k | B_k \rangle \otimes | R_k \rangle . \end{equation} where we've introduced bases for the remnant black hole state and the state of the radiation.  Now consider evolution of the system over time scales measured in units of the scrambling time of the remnant black hole.  Over such time scales it's reasonable to treat the radiation as freely propagating particles and neglect additional Hawking radiation.  The state will have evolved approximately to 
\begin{equation} | \psi \rangle = \sum a_k U_{kl} (t) | B_l \rangle \otimes U_{free} | R_k \rangle . \end{equation} Here $U_{kl}$ is a fairly random unitary matrix on the remnant black hole Hilbert space.  The systems are still entangled, but there is really no way for a detector ``measuring the radiation" to discern what the original old black hole state is at present.  This remark becomes even more sharp when one remembers that a complete quantum measurement consists of many different measurements of identically prepared systems in order to find out the correlation functions of non-commuting quantum operators.  There's no way that each of those measurements would find the same instant of time for $U_{kl} (t)$, even if one made the absurd postulate that one could always prepare the same $a_k$ at the Page time.  

Putting that remark aside, the main message of our paper is that these microscopic models of black hole evaporation show no sign of the famous ``black hole horizon" and the field theoretic divergences associated with it.  Instead they exhibit complex bound state structures, which when mapped into space-time have the size, energy and entropy relations of black holes, and decay into genuine particles with relativistic dispersion relations.  We should emphasize that the same is true of the most explicit picture of space-time locality in the AdS/CFT correspondence, the hyperbolic tensor network version of quantum error correcting codes.  In AdS/CFT a large black hole is a thermal state of the CFT.  In the micro-canonical ensemble this is the collection of states in some small energy window around a particular energy $E_0 > 1/R_{AdS}$.  These are all captured in the shell of the tensor network at a particular radius.  Shells at smaller radius miss most of the relevant states, while those at larger radius contain many that lie above the micro-canonical window.  Nothing singular is happening at this boundary.

Before proceeding, let us make clear that nothing in this note is meant to criticize the AdS/CFT resolution of the firewall paradox or of the notion of {\it non-isometric encoding} of information in quantum gravity\cite{akers} that has arisen from it\footnote{Indeed the Holographic Space-time/Hilbert Bundle formalism that the author has championed for years is, had he only known it, a form of non-isometric encoding.}.  All that is being challenged is the claim that anything in AdS/CFT has confirmed the bulk field theory model of the states that contribute to black hole entropy - some version of the ``brick wall" models of 't Hooft.  This is certainly not what happens in the two classes of models that are re-visited in this note.  

\section{The CGHS Model and Its Quantum Avatars}

The CGHS model is a particular form of two dimensional dilaton gravity coupled to $N$ massless fermions\footnote{In the original model the authors chose massless bosons, which are of course equivalent via bosonization.  We choose the fermion presentation to make contact with the underlying quantum models.} only through the metric.  It is the effective theory in the infinite throat of linear dilaton black holes in Type $II_B$ string theory.  In that context, the absence of coupling between the dilaton and the fermions comes about because they originate as modes of Ramond-Ramond fields.  The model is not supersymmetric because the black hole background breaks the SUSY of the underlying string theory.  The Lagrangian density  is
\begin{equation} {\cal L} = \frac{1}{2\pi} [ \sqrt{-g} [e^{-2\phi} (R + 4 (\nabla \phi)^2 + \frac{4}{L^2}) - \bar{\psi_i} e^m_a \gamma^a D_m \psi_i ] .\end{equation}

The pure gravity model has a one parameter set of static solutions, that take the form of black holes.  The model has a fixed length scale $L$, which we call the Planck length. The different black hole solutions can be characterized by their ADM energy and by the value of $S = e^{-2\phi}$ at the horizon, which as usual in dilaton graviton models we identify with entropy.  For a pulse consisting of $M$ fermions, with unit normalization, the ADM energy is of order $M/L$ and the entropy at the horizon is 
\begin{equation} S = \frac{M}{24} , \end{equation} which implies a Planck scale temperature, independent of the entropy.  For $M \neq 0$ the solution has a spacelike singularity, and the time for a geodesic to hit the singularity after horizon crossing is independent of $M$.  For $M = 0$, the singularity where $e^{-2\phi} = 0 $ is a null line.  This solution is called the linear dilaton vacuum.  

The authors of\cite{cghs} showed that arbitrary classical pulses of fermion energy, inserted into the linear dilaton vacuum, appeared to lead to the formation of a black hole. If the $N$ was large and a large number of fermions participated in the pulse, the entropy at the horizon was large and the authors speculated that the problem might be amenable to semi-classical analysis.  This proved incorrect.  The large $N$ semi-classical equations still had singularities, and despite years of work and a large number of papers\cite{papers} no real resolution of the problem of Hawking radiation in this model was found.

The resolution instead came from the convergence of a number of other strands of string theory research: large $N$ matrix models and two dimensional strings.  Bosonic string theory with two target space dimensions and a linear dilaton field that dials the tachyon mass to zero turned out to be an exactly soluble model on the world sheet. Moreover, it is also identical to a double scaled limit of matrix quantum mechanics, with a single matrix.  

If one gauges the $U(N)$ symmetry, the quantum mechanics of a single $N\times N$ Hermitian matrix with a single trace local action with potential  $\propto {\rm Tr} V(M)$ is exactly identical to the quantum mechanics of $N$ fermions in the same potential $\sum_i V(\lambda_i )$\cite{parisietal}  .  The large $N$ expansion of this model is a sum over the two dimensional topology of Feynman diagrams\cite{thooft}.  In the {\it double scaled limit}, these surfaces become smooth and we get a string theory, the bosonic string with $c = 1$ and a linear dilaton field to compensate for the ghost central charge. In terms of the matrix potential and the fermions, the double scaled limit corresponds to taking the fermi surface infinitesimally close to a quadratic maximum of the potential, and scaling parameters as $N \rightarrow \infty$ so that all but the quadratic term in the potential go to zero\cite{saclaygrossetc}.  Moore\cite{moore} provided an exact solution of the model in this limit.  Interestingly, string perturbation theory manages to agree with this solution, but only with extreme computational difficulty\cite{xiyin}.  The double scaled model is simply the non-relativistic field theory of fermions in the upside down oscillator potential. It has two parameters with the dimension of energy, the ``frequency", $\omega$ of the inverted oscillator, and the depth $d$ of the rescaled fermi surface below the maximum. 
$\omega/d$ plays the role of the string coupling.  When $d \gg \omega$ large, as we will see, perturbation theory is valid.   

The particle content of the $c = 1$ string is a single massless scalar/massless fermion.  There are vertex operators corresponding to dilaton and graviton and the worldsheet CFT can be considered the two dimensional linear dilaton vacuum solution of the Einstein equations, according to the usual connection between worldsheet beta functions and gravitational field equations.  To all orders in perturbation theory, one can consider the string scattering amplitudes as those of fermions scattering off the upside down oscillator potential starting from an asymptotic fermi surface far below the top of the potential.  

To see that we actually have relativistic fermions propagating we rewrite\cite{AKK}\cite{ms} (using units where $\omega = 1$)
 \begin{equation} (P_{\lambda} \pm \lambda) \equiv \sqrt{2} X_{\pm}  . \end{equation}
 $X_{\pm} $ Hermitian operators, which are canonically conjugate to each other and 
 \begin{equation} (P_{\lambda}^2 - \lambda^2)/2 = X_+ X_-  + i = X_- X_+ - i . \end{equation}  Viewed as differential operators, the product of two canonical conjugates is just $i \times$ the logarithmic derivative with respect to one of the coordinates.  Redefining the field so that the integration variable is that logarithmic coordinate, we see that the Hamiltonian is that of a left or right moving massless relativistic fermion.  When proper account is taken of the boundary conditions\cite{AKK}\footnote{A brief, very transparent account of this analysis can be found in the appendix to\cite{ms}.}, one sees that the model is exactly equivalent (for vanishing chemical potential, which corresponds to strong string coupling) to a pair of massless Dirac fermions propagating on a half line, with a scattering matrix that mixes them.  Lowering the fermi surface leads to a picture with less translation invariance.  The natural fields are the even and odd combinations under reflection symmetry.
 
 The appearance of {\it two species} of massless fermions implies that the perturbative equivalence with the bosonic string is misleading.  In fact, that corresponds to the unstable state in which the left fermi sea of non-relativistic fermions is completely empty\footnote{The instantons that indicate the instability in perturbative string theory are D brane instantons, ZZ branes\cite{ZZ}, which correspond to individual matrix eigenvalues tunneling over the barrier into the empty fermi sea.}. The real string model whose non-perturbative completion is Moore's fermionic field theory with finite fermi levels on both sides is the $\hat{c} = 1$ Type $0B$ string, with the two types of bosonized fermions being the RR and NS tachyons.  The fermi levels are chemical potentials for RR charge in the string model.  
 
 Apart from the detailed agreement\cite{xiyin} between perturbative string theory calculations and amplitudes in the fermionic field theory, the matching of vertex operators, the evidence that the non-relativistic fermionic field theory is really a general relativistic theory of two dimensional gravity coupled to fermions is hydrodynamic in nature.  We've seen the Minkowski background of the linear dilaton solution emerge from explicit calculation.  The dilaton background emerges similarly by computing the fermionic entropy of states restricted to lie at a fixed value of the Minkowski coordinate, which maps to a fixed value of the non-relativistic coordinate\cite{Das}.  This calculation gives the linear dilaton solution when we take the fermi level to the top of the potential.  The weakly coupled string regime is more complicated.
 
 One of the big disappointments of this model is that it does not exhibit the black hole solutions of linear dilaton gravity.  There are two obvious reasons why this is so.  The first is that the model is completely integrable and has an infinite number of conservation laws\cite{juanetal}, so cannot exhibit the chaotic dynamics expected of black holes.  The second is that, since we expect black hole formation to be non-perturbative in the string coupling, it must have to do with phenomena near the maximum of the potential.  However, even with strong coupling and fermi surfaces near the top of the potential, the entropy in the region near the top is small.
 
 We can solve the second problem by having a large number $N$ of fermion fields, which is to say we take $N$ copies of Moore's non-relativistic field theory.   This introduces even more unwanted symmetry, and to all orders in perturbation theory the copies don't talk to each other.  Recall that in the CGHS model one needed to have a large number of fermions to make the entropy of a dynamically formed black hole large. 
 
 We resolved the problem of too much symmetry by adding random quartic fermion couplings localized near the top of the potential to the model.  These couplings serve two purposes.  They break all the symmetries except those that correspond to space-time gauge symmetries in the interpretation of the model in terms of four dimensional charged black holes in Type II string theory.  And they are attractive.  Thus, if we send in a pulse containing a large number of different species of fermions, in wave-packets timed to arrive simultaneously near the top of the barrier, they will form a meta-stable self consistent bound state, with high entropy.  The length and energy scales involved are all determined by the curvature at the top of the potential, which we have called the Planck scale.  Since we're doing non-relativistic physics we can think of a self consistent Hartree potential with $M \sim N$ distinct species of fermions trapped in the $\geq 2M$ lowest meta-stable bound states.  This gives an entropy of order $M$.  The random couplings will move fermions around this state space on a time scale of order $\omega^{-1}$.  The meta-stable states will decay by emission of individual fermions at a rate of order $\omega$ so the ``Hawking temperature" is independent of the entropy.  And the time scale for a more or less free Dirac fermion to become enmeshed in this chaotic system is of order $\omega^{-1}$.   These are the qualitative properties of the linear dilaton black holes of the CGHS model.  The model is manifestly unitary and will of course have a normal Page curve.
 
 The main point of the present article is that, in this model, there is no appreciable distinction between the quantum state of an old black hole that has emitted half of its entropy and a newly formed one with the same entropy as the old one.  They're both typical meta-stable bound states of the same number of fermions in the same non-relativistic QM system undergoing the same chaotic dynamics.  The fermionic state in the old black hole is entangled with that of distant fermions that are propagating out towards null infinity.  The latter are described quite exactly by ordinary Poincare invariant QFT, but the ``region near the horizon" is not.  The non-relativistic four fermi interaction, whose form we have not specified completely since it's irrelevant to the argument, translates via the AKK dictionary into a complicated non-local four fermion interaction in terms of the Dirac fermions.  The transformation between the relativistic and non-relativistic position space variables guarantees that it's confined to a ``Planck scale" region of space.  Without the detailed mapping to a model that reproduces all orders in string perturbation theory one might be skeptical that this is a model of quantum gravity, even though it has states with the properties of the black hole solutions of linear dilaton gravity.  
 
 Like AdS/CFT, our models have no explicit gravitational field variables, and since we're in $1 + 1$ dimensional asymptotically flat space there are no fluctuations of boundary geometry or analogs of BMS charges.  We do have explicit variables describing fluctuations of the dilaton field in terms of coherent fluctuations of the fermi surface.  These have been discussed extensively in the literature\cite{polchinski}\cite{jevicki}.  
 
 \section{Matrix Models of Black Holes in 11 Dimensions}
 
 The BFSS matrix models of maximally supersymmetric gravity in $11 \geq d \geq 5$ start from the D(iscrete) L(ight) C(one) Q(uantized)\cite{sss} version of the theory.  For finite $N$ they describe a finite number of supergravitons scattering in Minkowski space, in the light front frame where dynamics resembles that of non-relativistic particles with variable mass.  Black holes have been understood in these models only in $8$ and $11$ dimensions.  We will concentrate on the $11$ dimensional case and discuss it only briefly since the main point is the same as the one we discovered in our two dimensional model.  
 
 The core of the black hole model is a classical solution of the large $N$ matrix equations of motion that behaves like an oscillating two dimensional membrane.  This solution exists because the matrix model is a ``fuzzification" of the light front Lagrangian for a supermembrane in 11D SUGRA.  We can then consider one block of the matrix to be evolving in this stationary solution and the interaction of other particles with it.  The particles are represented by other, small blacks of the matrix, and they interact by virtual exchange of off diagonal matrix elements.  In the D-brane origin of the model, these are open strings, and we called them "tethers", because it was shown\cite{bfssbh}\cite{hm} that they bind the supergravitons to the membrane.  Although individual supergravitons are bosons or fermions, different blocks are bound to different membrane sites, so they effectively have Boltzmann statistics.  The resulting meta-stable bound state has the size, entropy and energy relations of an eleven dimensional Schwarzschild black hole.  
 In\cite{matrixbhdecay} we showed that the state was indeed meta-stable.  It decays by ``snapping a tether", liberating an individual supergraviton.  The estimated rate of decay agrees with Hawking's formula up to order one constants.  A more recent black hole model\cite{ejmmatbh} in BFSS matrix models confirms similar estimates by different methods.  
 
 Once again we have a black hole model in terms of a (somewhat exotic) non-relativistic bound state, appearing as a legitimate state in a model with serious claims to be a valid model of quantum gravity\cite{bfssevidence}.  And again the model shows no sign of the field theoretic horizon and its associated divergence problems.  It does show a breakdown of perturbative scattering at a distance scale commensurate with where we expect the horizon to be.  Again the situation with respect to entanglement is the same as that for any run of the mill complex system that emits a lot of particles.  Just after the Page time, if the black hole plus radiation have been kept in complete isolation, the quantum state of the black hole could be retrieved from that of the radiation by a complete quantum measurement of the state of the radiation, which is in principle impossible.  Complete quantum measurements refer to multiple experiments on identically prepared systems.  One scrambling time of the old black hole after that, although the systems are still entangled, there is no way for even complete knowledge of the quantum state of the radiation at that later time to know what the current state of the black hole was without a detailed and correct model of the black hole dynamics.  And of course, if we've dropped a dust mote onto the black hole at any time during the process of evaporation or post Page time scrambling, the entanglement is destroyed. 
 
 One issue that was not explored in the early investigations of these black hole models in matrix theory was the question of infall time to the singularity.  Here, we are in for a surprise, because althought the asymptotic light front time of the model can be continued to a perfectly smooth coordinate across the future or past horizon of the black hole (but not both simultaneously) it is not at all the same thing as the proper time along infalling geodesics.  
 
 The metric of a boosted black hole with longitudinal momentum $P^+$ in Kerr-Schild coordinates is
 \begin{equation} ds^2 = - 2 dx^+ dx^- + d\vec{x}^2  + (\frac{r_s}{r})^{d-3} (\gamma dx^+ + dx^- \gamma )^2 . \end{equation}
 \begin{equation} \gamma \equiv (d - 2) A_{d-2} (L_P / r_s)^{d-3} L_P P^+ . \end{equation} 
  \begin{equation} r = \sqrt{1/2 (x^+ / \gamma - \gamma x^-)^2 + \vec{x}^2} . \end{equation} 
  
  For a radial geodesic that enters the horizon at zero transverse separation $\vec{x} = 0$ one finds that the infall time to the singularity is
  \begin{equation} \frac{\pi P^+}{4 (d - 2)(d-3) A_{d-2}} (\frac{L_P}{r_s})^{d-4}) L_P . \end{equation} The drastic difference between this and the conclusion that the proper infall time is of order $r_s$ for all $d$ is simply understood in terms of the transformation between coordinate and proper time.  This calculation resolves puzzles one would have otherwise had about why the matrix model of black holes did not exhibit the long proper time interval of non-dramatic evolution once external particles approach the horizon radius.  
 
 \section{Conclusions}
 
 In the author's opinion, the conclusion to be drawn from these examples {\it and} from the AdS/CFT resolution of the firewall paradox in terms of multiple CFT states that give the same correlation functions for ``simple" local measurements, is that the bulk EFT picture of the states near the horizon of a black hole is simply incorrect.  This is consistent with arguments\cite{tHooft}\cite{italians}\cite{CKN} that bulk EFT cannot account for the area law entropy of black holes.   Indeed, the {\it split} property of QFT shows that there cannot be QFT states like those postulated in the AdS/CFT resolution of the firewall paradox.
 
 Given a causal diamond $D$ and a region $R$ separated from $D$ by a space-like collar of finite size, the operator algebras of the two regions are contained in a tensor product $A(D) \otimes A(R)$ of Type I factors.  If we consider a density matrix entangling the Type I factors, which has entropy proportional to the maximal area on the diamond boundary in some units, then it's easy to prove, using the formula for the relative entropy of two states in terms of the expectation value of the stress tensor, and the ANEC bound on the stress tensor, that the expectation value of the stress tensor in the collar region cannot vanish.  Detailed estimates in free field theory\cite{tbpdk} show that many moments of the energy will be different than they are in the Hartle-Hawking state for a black hole.  
 
 In this author's opinion, the most promising model for the states on a black hole horizon is that proposed independently by Carlip\cite{carlip} and Solodukhin\cite{solo}.  This can be generalized to an arbitrary diamond\cite{BZ} and is supported by independent calculations for Ryu-Takayanagi diamonds\cite{deboeretal}\cite{VZ2} and for de Sitter space\cite{tbpdds}.  These authors modeled the horizon degrees of freedom as a $1 + 1$ dimensional conformal field theory (CFT) living on the ``stretched horizon" of the black hole. In\cite{BZ} it was shown that a cut-off CFT was sufficient to reproduce the results of calculations in GR, as long as the cutoff was above the scale where Cardy's formula for the smoothed eigenvalue distribution was valid.  In addition, the "stretched horizon" could be taken to be a small region in the vicinity of the maximal area surface on the diamond's boundary, generalizing the results of CS.  In\cite{hilbertbundles} the present author made a further conjecture about the nature of the boundary ``almost CFT", linking it to earlier work with Fischler\cite{hst}. 
 
 In the formalism described in those papers, the reduced density matrix describing an old black hole, a newly formed black hole, or a causal diamond in vacuum with the same area are identical.  The different systems differ by the way in which they relate to the larger system in which they're embedded, which for example could be an extremely large causal diamond in Minkowski space.  The density matrix is that of a boundary QFT whose variables are (in $4$ space-time dimensions) an $N \times (N + 1)$ matrix of cut-off $ 1 + 1 $ dimensional Dirac fields\footnote{The modular Hamiltonian is $L_0 + g^2 {\rm Tr} J_m J^m + const. $, where $J_m$ is the $U(N)$ current formed from the fields and $L_0$ the Virasoro generator.  We call this the empty diamond state $K_{E}(n)$ .} .   The larger diamond just has a larger matrix of fields.   In the Minkowski vacuum, the larger diamond is in its empty diamond state $K_E (T)$, $T \gg N$.  For the newly formed black hole $K_E (T)$ is modified by a projection operator, which sets the fields connecting those in the small matrix block to the rest of the fields to zero.  This is of order $2NT$ constraints.  For the old black hole we have of order $8 N T$ constraints, but the system consists of one large block of size $N$ and $(\sqrt{2} - 1)NT$ constraints that refer to small blocks of the matrix detached from the old black hole block.  
 
 If we look at pure quantum states, rather than reduced density matrices, then there is entanglement between the quantum state of the old black hole and the multi-block state of the small blocks, which is the holographic representation of the escaping Hawking radiation.  Notice that in this description we've completely neglected the fact that in a causal diamond of size $R$, the number of constraints is expected to change by of order $\pm L_P/R$, so that in actual fact all of these subsystems are becoming quantum mechanically entangled with tiny matrix blocks that represent soft gravitational radiation.  This happens whether there's a black hole in the process or not and becomes more and more pronounced as the center of mass scattering energy increases.  Sufficiently soft radiation is undetectable so in gravitational systems one is never in a practical position of being able to verify mathematical unitarity.  
 
 Thus the Carlip-Solodukhin ansatz for the behavior of black hole horizons exhibits the same lack of drama that our two explicit models did, and for the same reason.  {\it The QFT model of near horizon states is simply wrong.}  As shown by\cite{CKN} QFT has only been verified in situations that are compatible with the bounds of\cite{tHooft}\cite{CKN} on the number of QFT states in a finite sized causal diamond.  Over-reliance on it in the theory of black holes has led to more confusion than clarity.  
 
 To conclude, we should come back to the AdS/CFT resolution of the firewall paradox and ask whether there is anything in our models that corresponds to it.  The answer for the two dimensional model is a definitive no.  The Hilbert space at infinity is just the Fock space of relativistic fermions and the connection between them and the non-relativistic fermions liberated from the bound state is straight forward and more or less local far from the top of the potential and in the strong coupling regime where the connection to semi-classical dilaton gravity is most transparent.  In the BFSS matrix theory models, assuming that the large $N$ limit of those models at finite energy indeed reproduces the S matrix of M-theory in SUGRA Fock space, we can come to a similar conclusion if we're willing to be cavalier about taking the limit.  For any finite $N$, the process of black hole decay in light front frame is identical to the slow release of individual non-relativistic particles from a (somewhat exotic) meta-stable bound state.  The state of the leftover complex remnant is indeed entangled with that of the emitted particles (though in a way that is fairly useless for making conclusions about what's going on inside the bound state after any measurements on the radiation).  It's also true that many different states of the radiation will correspond to the same coarse grained properties of the remnant bound state.  What is not true is that the local EFT picture of the entropy of the black hole or the states near the ``black hole horizon" is a good approximation to the correct quantum state of the black hole in these models.  The AdS/CFT resolution of the firewall paradox does not establish the validity of the local EFT picture in AdS space either, although one gets the impression that some of the proponents of that resolution think that it does.  It actually says that the boundary encoding of local bulk physics has states that look nothing like any local EFT state in existence. No EFT states have entanglement between a local diamond and a remote space-like separated region, without affecting the values of stress tensor correlators in the region between the two.   This agrees with the conclusions of the two explicit models we have studied.  Those models begin from completely finite Lagrangians for matrix quantum mechanics and use well justified approximations in non-relativistic QM to establish the existence of meta-stable bound states with the scaling properties of black holes in asymptotically flat space in $2,11$ and $8$ dimensions.  The large $N$ limit of those models has been shown to reproduce a variety of perturbative string and supergravity scattering amplitudes in those same dimensions.  The states in the vicinity of the meta-stable bound state bear no resemblance to states near the horizon of a black hole in EFT and make no significant distinction between black holes that have evaporated past their Page time, and black holes that are newly formed.  The entanglement with distant radiation is a fragile, evanescent property that has no particular consequence for the local black hole wave function.

% =========================================================
\section*{Acknowledgments}
% =========================================================
I'd like to thank D. Harlow, Y. Zhao and A. Herderschee for conversations about the AdS/CFT resolution of the firewall paradox.  This paper was written while I was on sabbatical at the MIT Center for Theoretical Physics, a Leinweber Institute, whose hospitality is greatfully acknowledged.  

% =========================================================

\end{document}